\documentclass[aps,preprint,showpacs,superscriptaddress]{revtex4}  
\usepackage{graphicx}  
\usepackage[flushleft]{caption2} 
\usepackage{bm}        
\usepackage{citesort}
\usepackage{amssymb}   
\usepackage{amsmath}   

\begin{document}

\title{The Casimir force between parallel plates separated by anisotropic media }
\author{Gang \surname{Deng}}\email{deng@hust.edu.cn.}
\author{Bao-Hua \surname{Tan}} \email{tan_bh@126.com.}
\author{Ling \surname{Pei}}
\author{Ni \surname{Hu}}
\author{Jin-Rong \surname{Zhu}}
\affiliation{Hubei Collaborative Innovation Center for High-efficiency Utilization of Solar Energy, Hubei University of technology, Wuhan, 430068, P. R. China}
\affiliation{School of Science, Hubei University of Technology, Wuhan, 430068, P. R. China}

\begin{abstract}
The Casimir force between two parallel plates separated by anisotropic media is investigated. We theoretically calculate the Casimir force between two parallel plates when the interspace between the plates is filled with anisotropic media. Our result shows that the anisotropy of the material between the plates can significantly affect the Casimir force, especially the direction of the force. If ignoring the anisotropy of the in-between material makes the force to be repulsive (attractive), by contrast, taking the anisotropy into account may produce an extra attractive (repulsive) force. The physical explanation for this phenomenon is also discussed.

\end{abstract}


\maketitle

\section{Introduction}
The Casimir effect originating form the quantum fluctuations of the electromagnetic field is one of the most remarkable macroscopic effects of quantum physics \cite{1,2,3,4,5}. Great effort has been put into both theoretical and experimental studies on the Casimir effect\cite{4,5}, as it plays an important role in various fields of physics \cite{4}.

As we know, Lifshitz's original theory for the Casimir force is only applicable to isotropic dielectrics \cite{2,3}. However, anisotropy can bring us new features about the Casimir effect, like the Casimir torque. After the early works on the van der Waals interaction between anisotropic bodies in 1970s \cite{6,7}, Munday \emph{et al} numerically calculated the Casimir torque between parallel birefringent plates immersed in liquid, and proposed an experiment to observe this torque  \cite{8}. As supplemental works, Shao \emph{et al} gave the analytical expressions of the Casimir torque and repulsive Casimir force between two birefringent plates with constant permittivity and permeability \cite{9}. In one of our previous works, we calculated the Casimir torque between two parallel anisotropic plates with nontrivial permeabilities, and discussed the impact of magnetic properties of the plates on the torque \cite{10}. On the other hand, anisotropy can also affect the Casimir force directly. Romanowsky show that the orientation of the optical axis could have significant influence on Casimir force between highly anisotropic plates\cite{11}. In one of our previous works, we calculated the Casimir force between anisotropic metamaterial plates \cite{12}. The results show that the direction of Casimir force could change with both the separation and the anisotropy. Ran Zeng and his collaborators investigated the Casimir force between anisotropic single-negative metamaterial slabs and show that the electromagnetic responses of the metamaterial parallel and perpendicular to the optical axis affected the Casimir force differently \cite{13}. Recent years, the research works have extended to the interaction between anisotropic particles and a surface \cite{14,15}.

Most of the previous works \cite{7,8,9,10,11,12,13,14,15} were based on the assumption that the region between the two boundaries was vacuum or filled with isotropic media. However, in fact, this region can also be anisotropic. Some liquid, such as nitrobenzene and liquid crystal, can be anisotropic under special circumstances. One might naturally ask: what new phenomena can be seen if the region between the two slabs is filled with anisotropic media? Parsegian first calculated the van der Waals energy between two anisotropic bodies acting across a planar slab filled with a third anisotropic material in the non-retarded case \cite{6}. Kornilovitch investigated the Van der Waals interactions between flat surfaces in uniaxial anisotropic media in the non-retarded limit and discussed the effect of nonzero tilt between the optical axis and the surface normal on the interaction \cite{16}. On the other hand, repulsive Casimir force is always of great interests to the researchers \cite{2,3,4,5,8,9,12,13}. According to Lifshitz's theory, when the space between the two slabs is filled with isotropic media, the Casimir force can be repulsive if the permittivities of the plates ($\epsilon_{1}$ and $\epsilon_{2}$ ) and the interspatial media ($\epsilon_{3}$ ) satisfy the relation $\epsilon_{1}<\epsilon_{3}<\epsilon_{2}$ (or $\epsilon_{1}>\epsilon_{3}>\epsilon_{2}$)\cite{2,3}. However, if the interspatial media is birefringent, $\epsilon_{3}$ will have different values in different directions. Naturally, another question might be raised: when will the Casimir force be repulsive (or attractive) in this situation? Unfortunately, no clear clarification to this has been reported. Without doubt, the two problems above are of great importance. Therefore, it is necessary to study the possible new features of the Casimir effect between parallel plates separated by anisotropic media. In this work we use the quantized surface mode technique \cite{9,10,12,17} to calculate the Casimir force between two isotropic plates when the interspace between them is filled with anisotropic media. Our major concern is focused on the impact of the anisotropy on the Casimir force, especially the direction of the force. The result shows that the direction of the Casimir force can change with the anisotropy of the interspatial media. The physical understanding of this phenomenon is also investigated.  The detailed discussion will be presented in the following sections.

\section{The Casimir force between parallel plates separated by anisotropic media}

The system considered is shown in Fig. 1. Two isotropic plates (with the diameter $D$ and the thickness $d$) made of different materials are kept parallel to each other and separated with a distance $a$. The region between the plates is filled with uniaxial media. The plates and the media between them are considered to be nonmagnetic. The \emph{x-y} plane is chosen to be parallel to the surfaces of the plates. It should be noted that if we only consider the anisotropy of the media between the plates, but consider the plates to be isotropic,  there will be no Casimir torque. In this work, the optical axis of the anisotropic media is chosen to be in $z$ direction, which is often referred as the out-of-plane case.
\begin{figure}
  \includegraphics[width=0.3\textwidth]{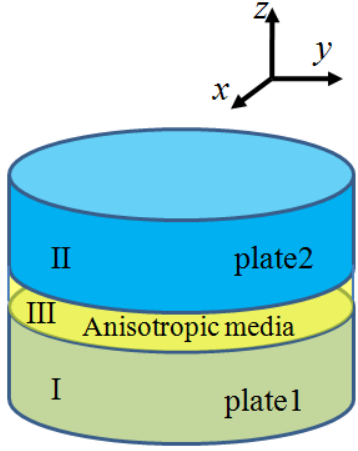}\\
  \renewcommand{\captionlabeldelim}{.}
  \caption{\label{fig:epsart}Two isotropic plates are separated by anisotropic media with a distance \emph{a}. The surfaces of the plates are parallel to the \emph{x}-\emph{y} plane. The optic axis of the interspatial anisotropic media is in \emph{z} direction.}
\end{figure}

 Assuming $D$ and $d$ are much greater than $a$, it is reasonable to disregard the edge effect and finite thickness effect. And the space can be approximately considered to be divided into three regions with corresponding permittivities, as shown in Fig.2. The relative permittivities can be expressed as diagonal matrixes, respectively, as following.
\begin{eqnarray}
    \epsilon_{1}=\left(
                    \begin{array}{ccc}
                      \epsilon_{1} & 0 & 0 \\
                      0 & \epsilon_{1} & 0 \\
                      0 & 0 & \epsilon_{1} \\
                    \end{array}
                  \right),\label{eqn1}\\
    \epsilon_{2}=\left(
                    \begin{array}{ccc}
                      \epsilon_{2} & 0 & 0 \\
                      0 & \epsilon_{2} & 0 \\
                      0 & 0 & \epsilon_{2} \\
                    \end{array}
                    \right),\label{eqn2}\\
    \epsilon_{3}=\left(
                    \begin{array}{ccc}
                      \epsilon_{3x} & 0 & 0 \\
                      0 & \epsilon_{3x} & 0 \\
                      0 & 0 & \epsilon_{3z} \\
                    \end{array}
                  \right),\label{eqn3}
\end{eqnarray}
where the subscripts $z$ and $x$ indicate the components parallel and perpendicular to the optical axis, respectively.
\begin{figure}
  \centering
  \includegraphics[width=0.3\textwidth]{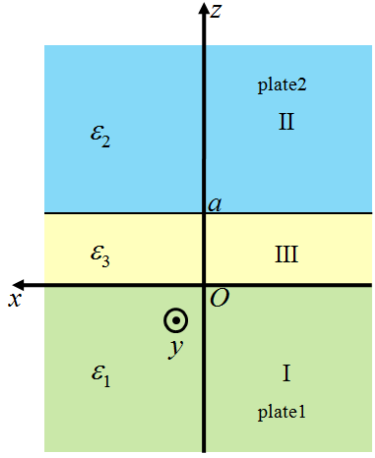}\\
  \renewcommand{\captionlabeldelim}{.}
  \caption{\label{fig:epsart}The schematic configuration. The space can be approximately considered to be divided into three regions. The optical property of each region is described by corresponding relative permittivity.}
\end{figure}

According to the quantized surface mode technique \cite{9,10,12,17}, we only need to consider the zero point energy associated with the surface modes \emph{q} which are exponentially decaying when \emph{z}$>$\emph{a} and \emph{z}$<$0. Because each region in Fig.2 is considered as homogenous, the electric and magnetic fields of the surface mode \emph{q} can be expressed as \cite{9,10,12,17,18}
\begin{eqnarray}
 \textbf{E}_{q}={\rm i} N [a_{q}\textbf{e}_{q}(\textbf{k})-a^{\dag}_{q}{\textbf{e}}^{*}_{q}(\textbf{k})],\label{eqn4}\\
 \textbf{H}_{q}=N[a_{q}\textbf{h}_{q}(\textbf{k})+a^{\dag}_{q}\textbf{h}^{*}_{q}(\textbf{k})],\label{eqn5}
\end{eqnarray}
where \emph{N} is the normalization factor, and the parameters $a_{q}$ and $a^{\dag}_{q}$ are the usual creation and annihilation operators, respectively. The parameters $\textbf{e}_{q}$ and $\textbf{h}_{q}$ are the electric and magnetic field polarization vectors. $\textbf{k}=(k_{x}, k_{y}, k_{z})$ is the wave vector.
We can choose the wave vector in \emph{x-y} plane to be parallel to the \emph{x} direction, and then the wave vector can be written as $\textbf{k}=(k_{x}, 0,k_{z})=K_{0}(\alpha, 0, \gamma)$, with $K_{0}=\omega c^{-1}$.

We introduce $M_{1}=\epsilon_{1}/\epsilon_{3x}$, $M_{2}=\epsilon_{2}/\epsilon_{3x}$ and $M_{3}=\epsilon_{3z}/\epsilon_{3x}$ to describe the relative values of $\epsilon_1$, $\epsilon_2$  and $\epsilon_{3z}$ to $\epsilon_{3x}$. $M_{3}$ can describe the anisotropy of the material filled between the plates. $M_{3}=1$ refers to isotropic. The Casimir energy per unit area at zero temperature can be expressed as ($\xi=-{\rm i}\omega$)
\begin{equation}
    E(\emph{a})=\frac{\hbar}{4\pi^{2}c^{2}}\int_{1}^{\infty}p{\rm d}p \int_{0}^{\infty}\epsilon_{3,\emph{x}} \xi^{2}{\rm d}\xi[\ln G_1({\rm i} \xi)+\ln G_2({\rm i} \xi)]
\label{eqn6}
\end{equation}
where \emph{c} is the speed of light in vacuum, and $\hbar$ is Plank constant divided by 2$\pi$. The variable $p$ is introduced as $p^{2}=1-\alpha^{2}/\epsilon_{3x}$.
The detailed representations for the functions $G_1({\rm i} \xi)$ and $G_2({\rm i} \xi)$ in Eq.(6) are expressed as (The detailed derivation of Eqs.(\ref{eqn6})-(\ref{eqn8}) is presented in Appendix A.)
\begin{eqnarray}
G_1({\rm i} \xi)=1-\frac{(s_{1}-p)(s_{2}-p)}{(s_{1}+p)(s_{2}+p)}{\rm exp}\left(-\frac{2pa\xi}{c}\sqrt{\epsilon_{3x}}\right),
\label{eqn7}\\
G_2({\rm i} \xi)=1-\frac{(s_{1}-M_{1}P)(s_{2}-M_{2}P)}{(s_{1}+M_{1}P)(s_{2}+M_{2}P)}{\rm exp}\left(-\frac{2Pa\xi}{c}\sqrt{\epsilon_{3x}}\right),
\label{eqn8}
\end{eqnarray}
where $s_{1,2}=\sqrt{M_{1,2}-1+p^2}$ and $P=\sqrt{(M_{3}-1+p^2)/M_{3}}$.
The Casimir force on the plates per unit area is
\begin{align}
    &F=\frac{\partial E(a)}{\partial a}=\frac{\hbar}{2\pi^{2}c^{3}}\int_{1}^{\infty}p{\rm d}p \int_{0}^{\infty}\epsilon_{3\emph{x}}^{3/2} \xi^{3}{\rm d}\xi\nonumber\\
    &\left[p\frac{1-G_{1}}{G_{1}}+P\frac{1-G_{2}}{G_{2}}\right]\nonumber\\
\label{eqn9}
\end{align}
As $G_2$ is a function of $M_3$, the Casimir force $F$ will also depend on $M_3$ which refers to the anisotropy of the material between the plates. And it can be found that $G_1$ does not depend on $M_3$, which means the anisotropy affects the force mainly through the second term in the brackets in Eq. (\ref{eqn9}).

For the case that the material between the plates is isotropic, $\epsilon_{3x}=\epsilon_{3z}=\epsilon_{3}$ ($M_{3}=1$ and $P=p$), Eq.(\ref{eqn9}) becomes

\begin{align}
    &F^{\rm isotropic}=\frac{\hbar}{2\pi^{2}c^{3}}\int_{1}^{\infty}p^{2}{\rm d}p \int_{0}^{\infty}\epsilon_{3}^{3/2} \xi^{3}{\rm d}\xi\nonumber\\
    &\left\{\left[\frac{s_{1}+p}{s_{1}-p}\frac{s_{2}+p}{s_{2}-p}{\rm exp}\left(\frac{2pa\xi}{c}\sqrt{\epsilon_{3}}\right)-1\right]^{-1}\right. \nonumber\\
    &\left.+\left[\frac{s_{1}+M_{1}p}{s_{1}-M_{1}p}\frac{s_{2}+M_{2}p}{s_{2}-M_{2}p}{\rm exp}\left(\frac{2pa\xi}{c}\sqrt{\epsilon_{3}}\right)-1\right]^{-1}\right\},
\label{eqn10}
\end{align}

which recovers Lifshitz's result about force on two bodies separated by a gap filled with a third isotropic media (equation 4.14 in Ref.\cite{3}).

Let's turn to the limiting case that the separation $a$ is larger than the characteristic absorption wavelength of the material. In this case, we can replace $\epsilon_{1}$,  $\epsilon_{2}$, $\epsilon_{3x}$ and $\epsilon_{3z}$ by their values at $\xi=0$, i.e. the static dielectric constants \cite{3}.
The approximate Casimir force can be expressed as following
\begin{equation}
   F\approx\frac{3\hbar c}{16\pi^2a^4\sqrt{\epsilon_{3x}}}\Psi
   \label{eqn11}
\end{equation}
with
\begin{equation}
 \Psi (M_1,M_2,M_3)=\int_{1}^{\infty}p{\rm d}p\left[\frac{1}{p^3}\left(\frac{s_{10}-p}{s_{10}+p}\right )\left(\frac{s_{20}-p}{s_{20}+p}\right )+\frac{1}{P^3}\left(\frac{s_{10}-M_1P}{s_{10}+M_1P}\right )\left(\frac{s_{20}-M_2P}{s_{20}+M_2P}\right )\right]
   \label{eqn12}
\end{equation}
where $s_{10}$ and $s_{20}$ are the values of $s_1$ and $s_2$ at $\xi=0$. (The detail of the approximate calculation of the integration in the Casimir force function is presented in Appendix B)

\section{The impact of the anisotropy of the media between the plates on the direction of the Casimir force}
The positive value of $F$ (or $\Psi$) in Eq. (\ref{eqn11}) corresponds to the attractive force, while the negative value corresponds to the repulsive force. As we are interested in the impact of the anisotropy of the interspatial media on the direction of the Casimir force, we can just discuss the sign of the function $\Psi$. From Eq.(\ref{eqn11}) and Eq.(\ref{eqn12}), it is clear that if the permittivities satisfy the relations $\epsilon_{1}<\epsilon_{3x}<\epsilon_{2}$ and $\epsilon_{1}<\epsilon_{3z}<\epsilon_{2}$ (or $\epsilon_{1}>\epsilon_{3x}>\epsilon_{2}$ and $\epsilon_{1}>\epsilon_{3z}>\epsilon_{2}$) at the same time, $\Psi$ will be minus and the force will be repulsive. This is because although $\epsilon_{3}$ has different values in different directions, all the possible values of $\epsilon_{3}$ are still in the range form $\epsilon_{2}$ to $\epsilon_{1}$. This can make the force repulsive, which is similar to the Lifshitz's result \cite{3}. However, what we mainly concern is not this case but the case when $\epsilon_{3x}$ or $\epsilon_{3z}$ is out of the range $(\epsilon_{2}, \epsilon_{1})$ (Here, $(\epsilon_{2}, \epsilon_{1})$ means the range form $\epsilon_{2}$ to $\epsilon_{1}$).

 First, we come to the case that only $\epsilon_{3x}$ is in the range $(\epsilon_{2}, \epsilon_{1})$. We let $\epsilon_{1}>\epsilon_{3x}>\epsilon_{2}$ ($M_1>1$ and $M_2<1$). Fig.3 shows the value of $\Psi$ for different $M_3$. The positive value of $\Psi$ corresponds to the attractive force, while the negative value corresponds to the repulsive force. Two vertical dash-dot lines, corresponding to $M_3=M_2$ and $M_3=M_1$, divide Fig.3 into 3 regions. In the middle region, where $M_2<M_3<M_1$ ($\epsilon_{2}<\epsilon_{3z}<\epsilon_{1}$), the force is always repulsive, which has been discussed previously. The other two regions (left and right) are the regions that we are interested in. In the left region, where $M_3<M_2$ ($\epsilon_{3z}<\epsilon_{2}$), the force is not always in the same direction and it can be either repulsive or attractive depending on the value of $M_3$ which refers to the anisotropy. In the region on the right, where $M_3>M_1$ ($\epsilon_{3z}>\epsilon_{1}$), the result is similar and the force can also be either repulsive or attractive depending on the value of $M_3$.

\begin{figure}
  \includegraphics[width=0.8\textwidth]{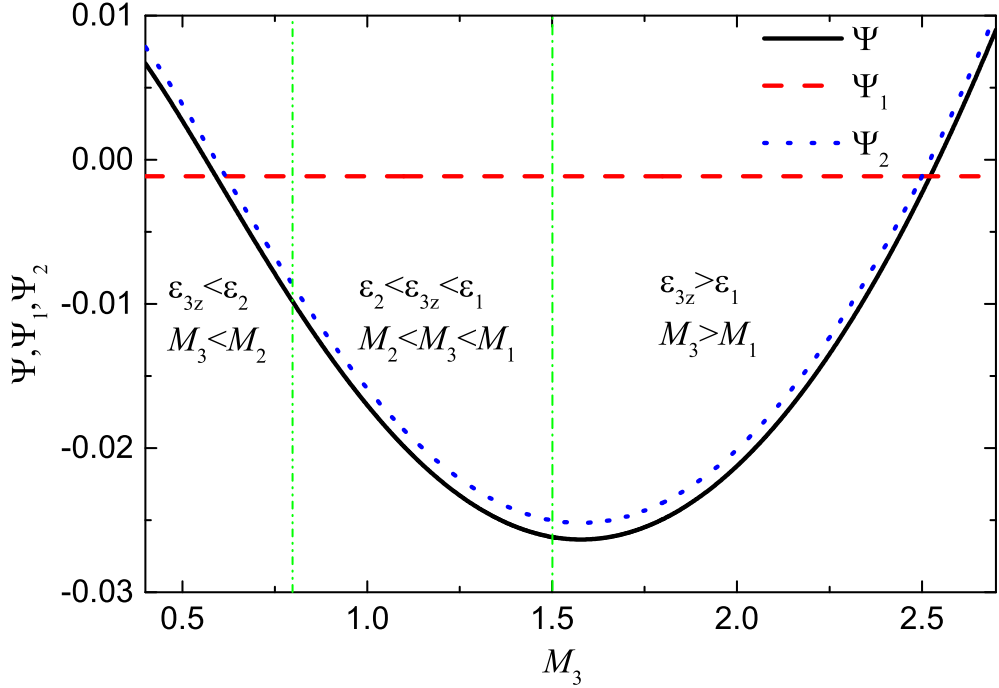}\\
  \renewcommand{\captionlabeldelim}{.}
  \caption{\label{fig:epsart}$\Psi$, $\Psi_1$ and $\Psi_2$ vs. $M_3$. $M_1=1.5$ and $M_2=0.8$. The two vertical dash-dot lines divide the range of $M_3$ into 3 parts corresponding to $\epsilon_{3z}<\epsilon_{2}$, $\epsilon_{2}<\epsilon_{3z}<\epsilon_{1}$, and $\epsilon_{3z}>\epsilon_{1}$ form left to right, respectively. The positive value corresponds to the attractive force, while the negative value corresponds to the repulsive force.}
\end{figure}

In order to see this more clearly, we can resolve $\Psi$ in Eq.(\ref{eqn12}) into two parts as flowing
\begin{equation}
\Psi(M_1,M_2,M_3)=\Psi_1+\Psi_2
\label{eqn13}
\end{equation}
with
\begin{eqnarray}
 \Psi_1(M_1,M_2)=\int_{1}^{\infty}p{\rm d}p\left[\frac{1}{p^3}\frac{s_{10}-p}{s_{10}+p}\frac{s_{20}-p}{s_{20}+p}\right]
   \label{eqn14}\\
 \Psi_2(M_1,M_2,M_3)=\int_{1}^{\infty}p{\rm d}p\left[\frac{1}{P^3}\frac{s_{10}-M_{10}P}{s_{10}+M_{10}P}\frac{s_{20}-M_{20}P}{s_{20}+M_{20}P}\right]
   \label{eqn15}
\end{eqnarray}

It is clear that $\Psi_1$ is independent of $\epsilon_{3z}$ and $M_3$, which means anisotropy of the in-between media does not affect $\Psi_1$. If $\epsilon_{1}>\epsilon_{3x}>\epsilon_{2}$ ($M_1>1$ and $M_2<1$), as we assumed previously, $\Psi_1$ will always be negative and contribute to a repulsive force, no matter how much the anisotropy ($M_3$) is. And $\Psi_2$ is the one that the anisotropy is mainly associated with. From Fig.3 we can see that $\Psi_2$ takes the dominant place in most of the range. It can be either positive or negative depending on the anisotropy. Although $\epsilon_{1}>\epsilon_{3x}>\epsilon_{2}$ always produces a repulsive force, the total force will not necessarily to be repulsive, if the anisotropy makes $\epsilon_{3z}$ to be out of this range, as shown in Fig.3. This is a result of the competition between $\Psi_1$ and $\Psi_2$. If the anisotropy makes $\Psi_2$ to be positive and have a greater amplitude than $\Psi_1$, the direction of the force will switch to attractive from repulsive. But without anisotropy, $\Psi_2$ will be surely negative and the force will be always repulsive. To achieve an attractive force, anisotropy is a must in this case.  Therefore, we can conclude that it is the anisotropy of the media between the plates that produces the attractive force in this case.

This can also be understood in an intelligible and vivid manner as flowing. According to Lifshitz's theory\cite{3}, the attractive force arises when $\epsilon_{3}$ is out of the range $(\epsilon_{2}, \epsilon_{1})$. However, in above case, $\epsilon_{3x}$ is assumed to be just in $(\epsilon_{2}, \epsilon_{1})$, which may always contribute to repulsive force. To produce an attractive force, one must let $\epsilon_{3z}$ to be a little far form this range. So that the "average value" (the quotation marks here mean that it is not the real mathematical average value \footnote{The quotation marks here mean that it is not the real mathematical average value. In Fact, it might be some value between $\epsilon_{3x}$ and $\epsilon_{3z}$. This is just a vivid way to make the problem easier to be understood. }) of the $\epsilon_{3}$  can be out of this range, and the attractive force can arise. And to make $\epsilon_{3z}$ out of the range $(\epsilon_{2}, \epsilon_{1})$, anisotropy is a must in this situation. This is why the attractive force ($\Psi>0$) happens only on the left and right edges of Fig.3.

\begin{figure}
  \includegraphics[width=0.8\textwidth]{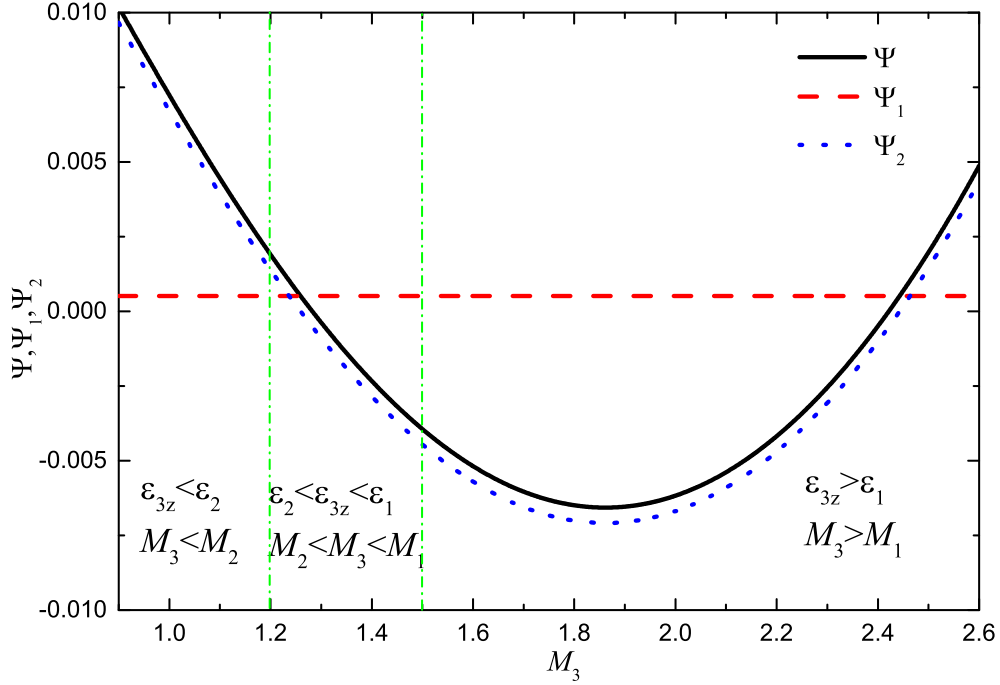}\\
  \renewcommand{\captionlabeldelim}{.}
  \caption{\label{fig:epsart}$\Psi$, $\Psi_1$ and $\Psi_2$ vs. $M_3$. $M_1=1.5$ and $M_2=1.1$. The two vertical dash-dot lines divide the range of $M_3$ into 3 parts corresponding to $\epsilon_{3z}<\epsilon_{2}$, $\epsilon_{2}<\epsilon_{3z}<\epsilon_{1}$, and $\epsilon_{3z}>\epsilon_{1}$ form left to right, respectively. The positive value corresponds to the attractive force, while the negative value corresponds to the repulsive force.}
\end{figure}
Now let's turn to the case that $\epsilon_{3x}$ is not in the range $(\epsilon_{2}, \epsilon_{1})$. If we let $\epsilon_{3x}<\epsilon_{2}$ ($M_1>M_2>1$), $\Psi_1$ will be always positive and produce an attractive force. As shown in Fig. 4,  $\Psi_2$ which is associated with anisotropy is still dominant in most of the range. Without anisotropy, $\Psi_2$ will be positive and the force will be always attractive. To achieve a repulsive force, $M_3$ must not be 0 in this case. We can also conclude that it is the anisotropy of the media between the plates that produces the repulsive force in this case. According to Lifshitz's theory \cite{3}, the force is repulsive only when $\epsilon_{1}>\epsilon_{3}>\epsilon_{2}$ (or $\epsilon_{1}<\epsilon_{3}<\epsilon_{2}$) is satisfied, and it is attractive in all the other cases. However, it can be found in Fig.4 that the total force can be repulsive, even when both $\epsilon_{3x}$ and $\epsilon_{3z}$ are out of the range $(\epsilon_{2}, \epsilon_{1})$ ! The explanation is similar to the previous case. As shown in Fig.4, if $\epsilon_{3x}<\epsilon_{2}$, to achieve a repulsive force, $M_3$ ($\epsilon_{3z}$) must be larger than $M_2$ ($\epsilon_{2}$) to bring the "average value" of $\epsilon_{3}$ in $(\epsilon_{2}, \epsilon_{1})$. However, it should not be too much greater than $M_1$ ($\epsilon_{1}$), as this may make the "average value" of $\epsilon_{3}$ beyond $\epsilon_{1}$. That is why the repulsive force only appears in the middle region of Fig.4.

From above discussion, we can find that the direction of the force is affected by $M_3$ which refers to the anisotropy of the media between the plates. If anisotropy of the in-between material makes the "average value" of  $\epsilon_{3}$ to be in the range $(\epsilon_{2}, \epsilon_{1})$, the force will be repulsive. Otherwise, the force will be attractive. The border curve defined by $\Psi=0$ is shown in Fig.5 (for convenient, $M_1$ is fixed to be 1.5). The regions with $\Psi>0$ are the attractive regions, while the regions with $\Psi<0$ are the repulsive regions.

\begin{figure}
  \includegraphics[width=0.8\textwidth]{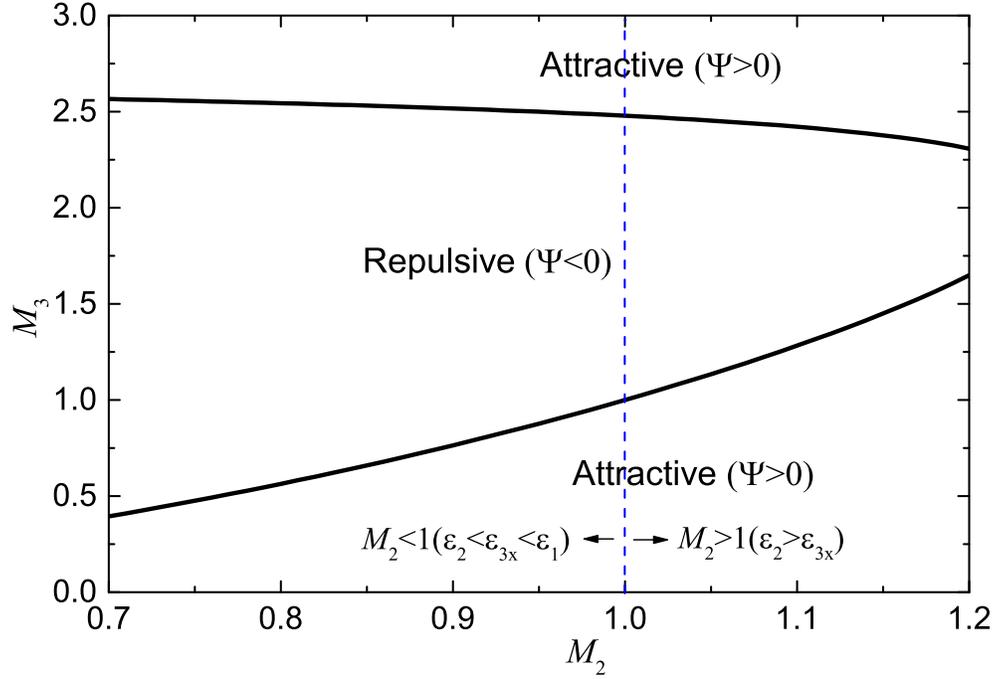}\\
  \renewcommand{\captionlabeldelim}{.}
  \caption{\label{fig:epsart}The repulsive and attractive regions in $M_2-M_3$ plane. The solid curve corresponds to $\Psi=0$ ($M_1=1.5$)}
\end{figure}

\section{Conclusion}
We have calculated the Casimir force between two plates separated by a gap filled with anisotropic media. The result shows that the anisotropy of the media between the plates plays an important, or sometimes dominant, role in the total force. The more important thing is that it can affect the direction of the force. If ignoring the anisotropy of the in-between material makes the force to be repulsive, by contrast, taking the anisotropy into account can produce an extra attractive force; and if ignoring the anisotropy of the in-between material makes the force to be attractive, taking the anisotropy into account can produce an extra repulsive force. This can be explained as the result of the competition of the attractive term and repulsive term in the force function. And it can also be understood, in an intelligible and vivid way, as the anisotropy makes the "average value" of the permittivity of the intermediate material to be in a certain range to produce the force in the corresponding direction. Finally, we should mention that it is still not easy to observe the above effects experimentally, since it is difficult to achieve high anisotropy in liquid.

\section*{Acknowledgment}
This work was partially supported by the Open Foundation of Hubei Collaborative Innovation Center for High-efficient Utilization of Solar Energy (HBSKFMS2014042 and HBSKFZD2014015), and the National Natural Science Foundation of China (11304091). The authors would like to show special thanks to Hubei University of Technology for providing early stage financial support form the Doctoral research program (BSQD12068)

\appendix

\section{The detailed derivation of the Casimir energy}
In this section we will present the detailed derivation of Eqs.(\ref{eqn6})-(\ref{eqn8}).
The electric and magnetic field polarization vectors $\textbf{e}_{q}$ and $\textbf{h}_{q}$ in Eqs.(\ref{eqn4}) and (\ref{eqn5}) with frequency $\omega$ can be determined according to the classical Maxwell equations (In SI Units)
\begin{eqnarray}
\nabla \times \textbf{E}=-\frac{\partial}{\partial t} \textbf{B}\label{eqna1}\\
\nabla \times \textbf{B}=\frac{1}{c^2}\frac{\partial}{\partial t} \epsilon_i \textbf{E}\label{eqna2}
\end{eqnarray}
$\epsilon_i$ (\emph{i}=1,2,3) is the relative permittivity in region \emph{i}, and its detailed representation can be found in Eqs.(\ref{eqn1})-(\ref{eqn3}), respectively.
The solution of the electric and magnetic fields can be expressed in the form of plane wave as
\begin{equation}
\left(
\begin{array}{c}
 \textbf{E}  \\
 \textbf{B}
\end{array}
\right)=\left(
\begin{array}{c}
 e_x  \\
 e_y \\
 e_z \\
 b_x  \\
 b_y \\
 b_z
\end{array}
\right){\rm e}^{{\rm i} K_{0}(\alpha x+\gamma z)-{\rm i}\omega t}
\label{eqna3a}
\end{equation}
$e_x$, $e_y$, and $e_z$ are the elements of the electric field (E) polarization vectors, while $b_x$, $b_y$, and $b_z$ are the elements of the magnetic field (B) polarization vectors. As we have assumed that all the materials considered in our work are nonmagnetic, we can have $b_x=\mu_0 h_x$, $b_y=\mu_0 h_y$, and $b_z=\mu_0 h_z$.

In Region I (as shown in Fig.2), substituting Eq.(\ref{eqn1}) and Eq.(\ref{eqna3a}) into Eqs.(\ref{eqna1}) and (\ref{eqna2}), we can have the eigenequations of the transverse elements of the electromagnetic field polarization vectors $e_x$, $e_y$, $b_x$, and $b_y$.
\begin{equation}
\left(
\begin{array}{cccc}
 0 & 0 & 1-\frac{\alpha ^2}{\epsilon_1} & 0  \\
 0 & 0 & 0  & -1 \\
 \epsilon_1 & 0 & 0 & 0  \\
 0 & -\epsilon_1+ \alpha ^2 & 0 & 0
\end{array}
\right)\left(
\begin{array}{c}
 e_x  \\
 e_y \\
 cb_y  \\
 cb_x
\end{array}
\right)=\gamma \left(\begin{array}{c}
 e_x  \\
 e_y \\
 cb_y  \\
 cb_x
\end{array}
\right)
\label{eqna3}
\end{equation}
The eigenvectors and the corresponding eigenvalues can be written in the matrix form as following:
\begin{equation}
W^{\rm I}=\left(
\begin{array}{cccc}
 0 & -\frac{{\rm i} t_1}{\epsilon_1} & 0 & \frac{{\rm i} t_1}{\epsilon_1}  \\
 -\frac{{\rm i}}{ t_1} & 0 & \frac{{\rm i}}{ t_1}  & 0 \\
 0& 1 & 0 & 1  \\
 1 & 0 & 1 & 0
\end{array}
\right)
\label{eqna4}
\end{equation}
\begin{equation}
\gamma ^{\rm I}=(
\begin{array}{cccc}
 \gamma _1^{\rm I} &
 \gamma _2^{\rm I} &
 \gamma _3^{\rm I} &
 \gamma _4^{\rm I}
\end{array}
)
=(
\begin{array}{cccc}
 -{\rm i}t_1 &
 -{\rm i}t_1 &
 {\rm i}t_1 &
 {\rm i}t_1
\end{array}
)
\label{eqna5}
\end{equation}
with $t^2_1=\alpha^2-\epsilon_1$. Each column of the matrix $W^{\rm I}$ represents an eigenvector. Four eigenvalues correspond to four independent mode solutions. The general solution of the electromagnetic field should be the linear superposition of the four mode solutions, and the superposition coefficients are the amplitudes of each mode. Then the transverse elements of the electromagnetic field in region I can be written as
 \begin{equation}
\left(\begin{array}{c}
 E_x  \\
 E_y \\
 cB_y  \\
 cB_x
\end{array}\right)
=W^{\rm I} \left(\begin{array}{c}
A^{\rm I}_1{\rm e}^{{\rm i} K_{0}(\alpha x+\gamma _1^{\rm I}z)} \\
 A^{\rm I}_2{\rm e}^{{\rm i} K_{0}(\alpha x+\gamma _2^{\rm I}z)} \\
 A^{\rm I}_3{\rm e}^{{\rm i} K_{0}(\alpha x+\gamma _3^{\rm I}z)}  \\
 A^{\rm I}_4{\rm e}^{{\rm i} K_{0}(\alpha x+\gamma _4^{\rm I}z)}
\end{array}\right){\rm e}^{-{\rm i}\omega t}
\label{eqna6}
\end{equation}
where $A^{\rm I}_j$ (\emph{j}=1,2,3,4) is the amplitude of the \emph{j}th mode solution.

Similarly, in region II, where it is also isotropic, the transverse elements of the electromagnetic field can be written as
  \begin{equation}
\left(\begin{array}{c}
 E_x  \\
 E_y \\
 cB_y  \\
 cB_x
\end{array}\right)
=W^{\rm II} \left(\begin{array}{c}
A^{\rm II}_1{\rm e}^{{\rm i} K_{0}(\alpha x+\gamma _1^{\rm II}z)} \\
 A^{\rm II}_2{\rm e}^{{\rm i} K_{0}(\alpha x+\gamma _2^{\rm II}z)} \\
 A^{\rm II}_3{\rm e}^{{\rm i} K_{0}(\alpha x+\gamma _3^{\rm II}z)}  \\
 A^{\rm II}_4{\rm e}^{{\rm i} K_{0}(\alpha x+\gamma _4^{\rm II}z)}
\end{array}\right){\rm e}^{-{\rm i}\omega t}
\label{eqna7}
\end{equation}
with
\begin{equation}
W^{\rm II}=\left(
\begin{array}{cccc}
 0 & -\frac{{\rm i} t_2}{\epsilon_2} & 0 & \frac{{\rm i} t_2}{\epsilon_2}  \\
 -\frac{{\rm i}}{ t_2} & 0 & \frac{{\rm i}}{ t_2}  & 0 \\
 0& 1 & 0 & 1  \\
 1 & 0 & 1 & 0
\end{array}
\right)
\label{eqna8}
\end{equation}
and
\begin{equation}
\gamma ^{\rm I}=(
\begin{array}{cccc}
 \gamma _1^{\rm II} &
 \gamma _2^{\rm II} &
 \gamma _3^{\rm II} &
 \gamma _4^{\rm II}
\end{array}
)
=(
\begin{array}{cccc}
 -{\rm i}t_2 &
 -{\rm i}t_2 &
 {\rm i}t_2 &
 {\rm i}t_2
\end{array}
)
\label{eqna9}
\end{equation}
where $t^2_2=\alpha^2-\epsilon_2$, and $A^{\rm II}_j$ (\emph{j}=1,2,3,4) is the amplitude of the \emph{j}th mode solution.

In region III, where it is anisotropic, the case is different, as the permittivity has different form as shown in Eq.(\ref{eqn3}).
Substitute Eq.(\ref{eqn3}) and Eq.(\ref{eqna3a}) into Eqs.(\ref{eqna1}) and (\ref{eqna2}), and we can have the eigenequations of the transverse elements of the electromagnetic field polarization vectors $ e_x$, $e_y$, $b_x$, and $b_y$.
\begin{equation}
\left(
\begin{array}{cccc}
 0 & 0 & 1-\frac{\alpha ^2}{\epsilon_{3z}} & 0  \\
 0 & 0 & 0  & -1 \\
 \epsilon_{3x} & 0 & 0 & 0  \\
 0 & -\epsilon_{3x}+ \alpha ^2 & 0 & 0
\end{array}
\right)\left(
\begin{array}{c}
 e_x  \\
 e_y \\
 cb_y  \\
 cb_x
\end{array}
\right)=\gamma \left(\begin{array}{c}
 e_x  \\
 e_y \\
 cb_y  \\
 cb_x
\end{array}
\right)
\label{eqna10}
\end{equation}

The eigenvectors and the corresponding eigenvalues can be written in the matrix form as following:
\begin{equation}
W^{\rm III}=\left(
\begin{array}{cccc}
 0 & -\frac{{\rm i} t_{3z}}{\epsilon_{3x}} & 0 & \frac{{\rm i} t_{3z}}{\epsilon_{3x}}  \\
 -\frac{{\rm i}}{ t_{3x}} & 0 & \frac{{\rm i}}{ t_{3x}}  & 0 \\
 0& 1 & 0 & 1  \\
 1 & 0 & 1 & 0
\end{array}
\right)
\label{eqna11}
\end{equation}
\begin{equation}
\gamma ^{\rm III}=(
\begin{array}{cccc}
 \gamma _1^{\rm III} &
 \gamma _2^{\rm III} &
 \gamma _3^{\rm III} &
 \gamma _4^{\rm III}
\end{array}
)
=(
\begin{array}{cccc}
 -{\rm i}t_{3x} &
 -{\rm i}t_{3z} &
 {\rm i}t_{3x} &
 {\rm i}t_{3z}
\end{array}
)
\label{eqna12}
\end{equation}
with $t^2_{3x}=\alpha^2-\epsilon_{3x}$ and $t^2_{3z}=(\alpha^2-\epsilon_{3z})(\epsilon_{3x} / \epsilon_{3z})$ .
Similarly, in region III, the transverse elements of the electromagnetic field can be written as
 \begin{equation}
\left(\begin{array}{c}
 E_x  \\
 E_y \\
 cB_y  \\
 cB_x
\end{array}\right)
=W^{\rm III} \left(\begin{array}{c}
A^{\rm III}_1{\rm e}^{{\rm i} K_{0}(\alpha x+\gamma _1^{\rm III}z)} \\
 A^{\rm III}_2{\rm e}^{{\rm i} K_{0}(\alpha x+\gamma _2^{\rm III}z)} \\
 A^{\rm III}_3{\rm e}^{{\rm i} K_{0}(\alpha x+\gamma _3^{\rm III}z)}  \\
 A^{\rm III}_4{\rm e}^{{\rm i} K_{0}(\alpha x+\gamma _4^{\rm III}z)}
\end{array}\right){\rm e}^{-{\rm i}\omega t}
\label{eqna13}
\end{equation}
where $A^{\rm III}_j$ (\emph{j}=1,2,3,4) is the amplitude of the \emph{j}th mode solution.

As the surface modes should be exponentially decaying for $z>0$ and $z<a$,  we have $A^{\rm I}_3=A^{\rm I}_4 =A^{\rm II}_1=A^{\rm II}_2=0$ \cite{9,17}. As the transverse elements of the electromagnetic field are continuous at $z=0$, we have
 \begin{equation}
\left(\begin{array}{cc}
 0 &  -\frac{{\rm i} t_1}{\epsilon_1} \\
 -\frac{{\rm i}}{t_1} & 0\\
 0 & 1  \\
 1 & 0
\end{array}\right)\left(\begin{array}{c}
A^{\rm I}_1 \\
 A^{\rm I}_2
\end{array}\right)
=W^{\rm III} \left(\begin{array}{c}
A^{\rm III}_1 \\
 A^{\rm III}_2 \\
 A^{\rm III}_3  \\
 A^{\rm III}_4
\end{array}\right)
\label{eqna14}
\end{equation}

As the transverse elements of the electromagnetic field are continuous at $z=a$, we have,
 \begin{equation}
\left(\begin{array}{cc}
 0 &  \frac{{\rm i} t_2}{\epsilon_2} \\
 \frac{{\rm i}}{t_2} & 0\\
 0 & 1  \\
 1 & 0
\end{array}\right)\left(\begin{array}{c}
A^{\rm II}_3 {\rm e}^{-K_0 t_2 a}\\
 A^{\rm II}_4 {\rm e}^{-K_0 t_2 a}
\end{array}\right)
=W^{\rm III} \left(\begin{array}{c}
A^{\rm III}_1 {\rm e}^{K_0 t_{3x} a}\\
 A^{\rm III}_2 {\rm e}^{K_0 t_{3z} a}\\
 A^{\rm III}_3 {\rm e}^{-K_0 t_{3x} a} \\
 A^{\rm III}_4 {\rm e}^{-K_0 t_{3z} a}
\end{array}\right)
\label{eqna15}
\end{equation}

Eqs.(\ref{eqna14})and (\ref{eqna15}) include eight linear homogeneous equations relating the unknown parameters $A^{\rm I}_1$, $A^{\rm I}_2$, $A^{\rm II}_3$, $A^{\rm II}_4$, $A^{\rm III}_1$, $A^{\rm III}_2$, $A^{\rm III}_3$, and $A^{\rm III}_4$. It has the nontrivial solutions if the determinant of its coefficients is equal to zero, which leads to the equation for the determination of the proper frequency $\omega$ . And the equation of determinant of coefficients \footnote{In fact, we cancel $A^{\rm III}_1$, $A^{\rm III}_2$, $A^{\rm III}_3$, and $A^{\rm III}_4$, and reduce the problem to be 4 linear equations with 4 unknown parameters $A^{\rm I}_1$, $A^{\rm I}_2$, $A^{\rm II}_3$ and $A^{\rm II}_4$.  This makes the determinant of coefficients to be $4\times 4$ rather than $8\times 8$.} equalling zero can be transformed into following form
\begin{equation}
Y*[G_1(\omega _{\perp})*G_2(\omega _{\parallel})]=0
\label{eqna16}
\end{equation}
where \emph{Y} is a function that is not always equal to zero. The subscripts $\perp$ and $\parallel$ indicate the mode with the polarization of the electric field perpendicular and parallel to the plane formed by $\vec{k}_{\parallel}=(k_{x}, k_{y})$ and \emph{z} (As we have assumed $\vec{k}_{\parallel}$ to be parallel to the \emph{x} direction, this plane is also \emph{x-z} plane). Introducing $\omega = {\rm i} \xi$, the functions $G_1(\omega _{\perp})$ and $G_2(\omega _{\parallel})$ can be expressed as following (Using the relation $p^{2}=1-\alpha^{2}/\epsilon_{3x}$, we can transform $t^2_1$, $t^2_2$, $t^2_{3x}$ and  $t^2_{3z}$ into the functions of \emph{p} or $s_{1,2}$ \footnote {With the relation $p^{2}=1-\alpha^{2}/\epsilon_{3x}$, we have $t^2_1=-s^2_1\epsilon_{3x}$, $t^2_2=-s^2_2\epsilon_{3x}$, $t^2_{3x}=-p^2\epsilon_{3x}$, and $t^2_{3z}=[\epsilon_{3x}(1-p^2)-\epsilon_{3z}]\frac{\epsilon_{3x}}{\epsilon_{3z}}$. This transformation might make the variables similar to the notations used in Lifshitz's theory.})

\begin{eqnarray}
G_1(\omega _{\perp})=G_1({\rm i} \xi)=1-\frac{(s_{1}-p)(s_{2}-p)}{(s_{1}+p)(s_{2}+p)}{\rm exp}\left(-\frac{2pa\xi}{c}\sqrt{\epsilon_{3x}}\right),
\label{eqna17}\\
G_2(\omega _{\parallel})=G_2({\rm i} \xi)=1-\frac{(\epsilon_{3x} s_{1}-\epsilon_1 \sqrt{\frac{\epsilon_{3x}(p^2-1)+\epsilon_{3z}}{\epsilon_{3z}}})(\epsilon_{3x} s_{2}-\epsilon_2 \sqrt{\frac{\epsilon_{3x}(p^2-1)+\epsilon_{3z}}{\epsilon_{3z}}})}{(\epsilon_{3x} s_{1}+\epsilon_1 \sqrt{\frac{\epsilon_{3x}(p^2-1)+\epsilon_{3z}}{\epsilon_{3z}}})(\epsilon_{3x} s_{2}+\epsilon_2 \sqrt{\frac{\epsilon_{3x}(p^2-1)+\epsilon_{3z}}{\epsilon_{3z}}})}{\rm e}^{-\frac{2a\xi\sqrt{\epsilon_{3x}^2(p^2-1)+\epsilon_{3z} \epsilon_{3x}}}{c \sqrt{\epsilon_{3z}}}},
\label{eqna18}
\end{eqnarray}
Eq.(\ref{eqna17}) is the same as Eq.(\ref{eqn7}). Substituting $M_{1}=\epsilon_{1}/\epsilon_{3x}$, $M_{2}=\epsilon_{2}/\epsilon_{3x}$, $M_{3}=\epsilon_{3z}/\epsilon_{3x}$ and $P=\sqrt{(M_{3}-1+p^2)/M_{3}}$ into Eq.(\ref{eqna18}), we will get Eq.(\ref{eqn8})

The Casimir energy can be expressed as \cite{9,17}
\begin{equation}
    E(\emph{a})=\frac{\hbar}{8\pi^{2}}\int k{\rm d}k \int_{0}^{2\pi}{\rm d}\theta \left(\sum\limits_{n}\omega _{n, \perp}+\sum\limits_{n}\omega _{n, \parallel}
\right)
\label{eqna19}
\end{equation}
The summations over \emph{n} can be performed with the help of the argument theorem which has been applied in Refs.\cite{4,9,17}. And then we have
\begin{equation}
    E(\emph{a})=\frac{\hbar}{8\pi^{3}}\int k{\rm d}k \int_{0}^{2\pi}{\rm d}\theta \int_{0}^{\infty} (\ln G_1+\ln G_2){\rm d}\xi
\label{eqna20}
\end{equation}

Substituting $\alpha ^2 = \epsilon _{3x}(1-p^2)$ and $k=\frac{\omega}{c}\alpha$ into Eq.(\ref{eqna20}), we can arrive at Eq.(\ref{eqn6}).

\section{The approximate calculation of the integration in the Casimir force}
In this part we will introduce how we approximately calculate the integration over $\xi$ in the Casimir force in the limiting case.
The Casimir force in Eq.(\ref{eqn9}) can be rewritten as following
\begin{equation}
F=F_1+F_2
\label{eqnb1}
\end{equation}
with
\begin{eqnarray}
F_{1}=\frac{\hbar}{2\pi^{2}c^{3}}\int_{1}^{\infty}p^2{\rm d}p\int_{0}^{\infty}\epsilon_{3x}^{3/2} \xi^{3}{\rm d}\xi\left[\frac{1-G_{1}}{G_{1}}\right]
\label{eqnb2}\\
F_{2}=\frac{\hbar}{2\pi^{2}c^{3}}\int_{1}^{\infty}pP{\rm d}p\int_{0}^{\infty}\epsilon_{3x}^{3/2} \xi^{3}{\rm d}\xi\left[\frac{1-G_{2}}{G_{2}}\right]
\label{eqnb3}
\end{eqnarray}
The detailed expressions of $G_1$ and $G_2$ can be found in Eq. (\ref{eqn7}) and Eq. (\ref{eqn8}).

For the limiting case that the separation $a$ is larger than the characteristic absorption wavelength of the material, we can replace $\epsilon_{1}$,  $\epsilon_{2}$, $\epsilon_{3x}$ and $\epsilon_{3z}$ by their values at $\xi=0$, i.e. the statistic dielectric constants \cite{3}. We introduce Eq.(\ref{eqnb2}) a new variable of integration $X=2pa\xi\sqrt{\epsilon_{3x}}/c$, and the integration should be taken over $X$ and $p$. Eq.(\ref{eqnb2}) becomes
\begin{align}
 F_{1}=&\frac{\hbar c}{32\pi^2a^4\sqrt{\epsilon_{3x}}}\int_{1}^{\infty}\frac{{\rm d}p}{p^2}\int_{0}^{\infty}X^{3}{\rm d}X\nonumber\\
 &\left[\frac{s_{10}+p}{s_{10}-p}\frac{s_{20}+p}{s_{20}-p}{\rm exp}(X)-1\right]^{-1}
\label{eqnb4}
\end{align}
The integration over $X$ can be taken by using approximate equation $\frac{m}{n!}\int_{0}^{\infty}\frac{x^{n}{\rm d}x}{m{\rm exp}(x)-1}\approx 1$\cite{3}.
\begin{equation}
   F_{1}\approx\frac{3\hbar c}{16\pi^2a^4\sqrt{\epsilon_{3x}}}\int_{1}^{\infty}\frac{{\rm d}p}{p^2}\left[\frac{s_{10}-p}{s_{10}+p}\frac{s_{20}-p}{s_{20}+p}\right]
   \label{eqnb5}
\end{equation}
similarly, we can have the approximate result of $F_{2}$
\begin{equation}
   F_{2}\approx\frac{3\hbar c}{16\pi^2a^4\sqrt{\epsilon_{3x}}}\int_{1}^{\infty}\frac{p{\rm d}p}{P^3}\left[\frac{s_{10}-M_{1}P}{s_{10}+M_{1}P}\frac{s_{20}-M_{2}P}{s_{20}+M_{2}P}\right]
   \label{eqnb6}
\end{equation}
Then we can have the result in Eq. (\ref{eqn11}). And from Eqs. (\ref{eqn14}) and (\ref{eqn15}) we can find that
 \begin{eqnarray}
   F_{1}=\frac{3\hbar c}{16\pi^2a^4\sqrt{\epsilon_{3x}}}\Psi_1
   \label{eqnb7}\\
   F_{2}=\frac{3\hbar c}{16\pi^2a^4\sqrt{\epsilon_{3x}}}\Psi_2
   \label{eqnb8}
\end{eqnarray}


\begin{thebibliography}{99}
\bibitem{1} H. B. G. Casimir, Proc. K. Ned. Akad. Wet. \textbf{51}, 793 (1948).
\bibitem{2} E. M. Lifshitz, Sov. Phys. JETP \textbf{2}, 73 (1956).
\bibitem{3} I. E.Dzyaloshinskii, E. M. Lifshitz, and  L. P. Pitaevskii, Adv. Phys. \textbf{10}, 165(1961).
\bibitem{4} M. Bordag, U. Mohideen, and V. M. Mostepanenko, Phys. Rep. \textbf{353}, 1 (2001).
\bibitem{5} G. L. Klimchitskaya, U. Mohideen, and V. M. Mostepanenko, Reviews of Modern Physics, \textbf{81}, 1827 (2009).
\bibitem{6} V. A. Parsegian, and George H. Weiss, J. Adhes. \textbf{3}, 259 (1972).
\bibitem{7} Yu S. Barash, Radiophys. Quant. Elec. \textbf{21}, 1138 (1978).
\bibitem{8} Jeremy N. Munday, Davide Iannuzzi, Yuri Barash, and Federico Capasso, Phys. Rev. A, \textbf{71}, 042102 (2005).
\bibitem{9} C. G. Shao, A. H. Tong, and J. Luo, Phys. Rev. A, \textbf{72}, 022102 (2005); C. G. Shao, D. L. Zheng and J. Luo, Phys. Rev. A, \textbf{74} 012103 (2006).
\bibitem{10}Gang Deng, Zhong-Zhu Liu, and Jun Luo, Phys. Rev. A, \textbf{80}, 062104 (2009)
\bibitem{11}Mark B. Romanowsky and Federico Capasso, Phys. Rev. A, \textbf{78}, 042110 (2008).
\bibitem{12}G. Deng, Z. Z. Liu and J. Luo, Phys. Rev. A, \textbf{78}, 062111 (2008).
\bibitem{13}Ran Zeng, Yaping Yang, and Shiyao Zhu, Phys. Rev. A, \textbf{87} 063823 (2013).
\bibitem{14}K. V. Shajesh and M. Schaden, Phys. Rev. A, \textbf{85}, 012523 (2012).
\bibitem{15}T. Taillandier-Loize, J. Baudon, G. Dutier, F. Perales, M. Boustimi, and M. Ducloy, Phys. Rev. A, \textbf{89} 052514 (2014).
\bibitem{16}Pavel E. Kornilovitch, J. Phys.: Condens. Matter, \textbf{25}, 035102(2013).
\bibitem{17}Fei Zhou, and Larry Spruch, Phys. Rev. A 52, 297 (1995)
\bibitem{18}P. W. Milonni, \emph{The quantum Vacuum} (Academic Press, Boston, 1994).
\end{thebibliography}
\end{document}